\documentclass[12pt,preprint,xdvi]{aastex}

\usepackage{natbib}
\usepackage{graphicx}
\usepackage{times}
\usepackage{natbib}
\usepackage{graphicx}

\newcommand{\be}{\begin{equation}}
\newcommand{\ee}{\end{equation}}
\newcommand\eq{eq.}

\newcommand\fig{Fig.}

\def\va{v_{\rm A}}
\def\vae{v_{_{\rm A},e}}
\def\ta{T^{\rm A}_e}

\def\vvec{{\bf v}}
\def\uvec{{\bf u}}
\def\vpar{v_{_{\parallel}}}
\def\xvec{{\bf x}}
\def\that{{\bf \hat{t}}}
\def\xhat{{\bf \hat{x}}}

\def\zhat{{\bf \hat{z}}}
\def\momvec{{\bf P}}

\def\quarter{\hbox{${1\over4}$}}
\def\disc#1{{[\![#1]\!]}}
\def\avg#1{{\bigl\langle#1\bigr\rangle}}

\begin{document}

\title{Gas-dynamic shock heating of post-flare loops due to 
retraction following localized, impulsive reconnection}

\author{D.W. Longcope,$^1$ S.E. Guidoni,$^1$ and M.G. Linton$^2$}
\affil{1. Department of Physics, Montana State University,
  Bozeman, Montana 59717\\
2. Naval Research Laboratory, Washington, D.C.}

\keywords{MHD --- shock waves --- Sun: flares}


\begin{abstract}
We present a novel model in which field lines shortening after
localized, three-dimensional reconnection heat the
plasma as they compress it.  The shortening progresses
away from the reconnection site at the Alfv\'en speed, 
releasing magnetic energy and generating parallel, compressive flows.
These flows, which are highly supersonic when $\beta\ll1$,
collide in a pair of strong gas-dynamic shocks at which 
both the mass density and temperature are raised.
Reconnecting field lines initially 
differing by more that $100^{\circ}$ can produce
a concentrated knot of plasma hotter that $20$ MK at the loop's apex, 
consistent with observations.
In spite of these high temperatures, the shocks convert
less than $10\%$ of the liberated magnetic energy into heat --- the
rest remains as kinetic energy of bulk motion.  These gas-dynamic
shocks arise only when the reconnection is impulsive and localized in
all three dimensions; they are distinct from the slow
magnetosonic shocks of the Petschek steady-state reconnection model.
\end{abstract}

\date{Draft: \today}

\section{Introduction}

\begin{figure}[htp]
\epsscale{1.0}
\plotone{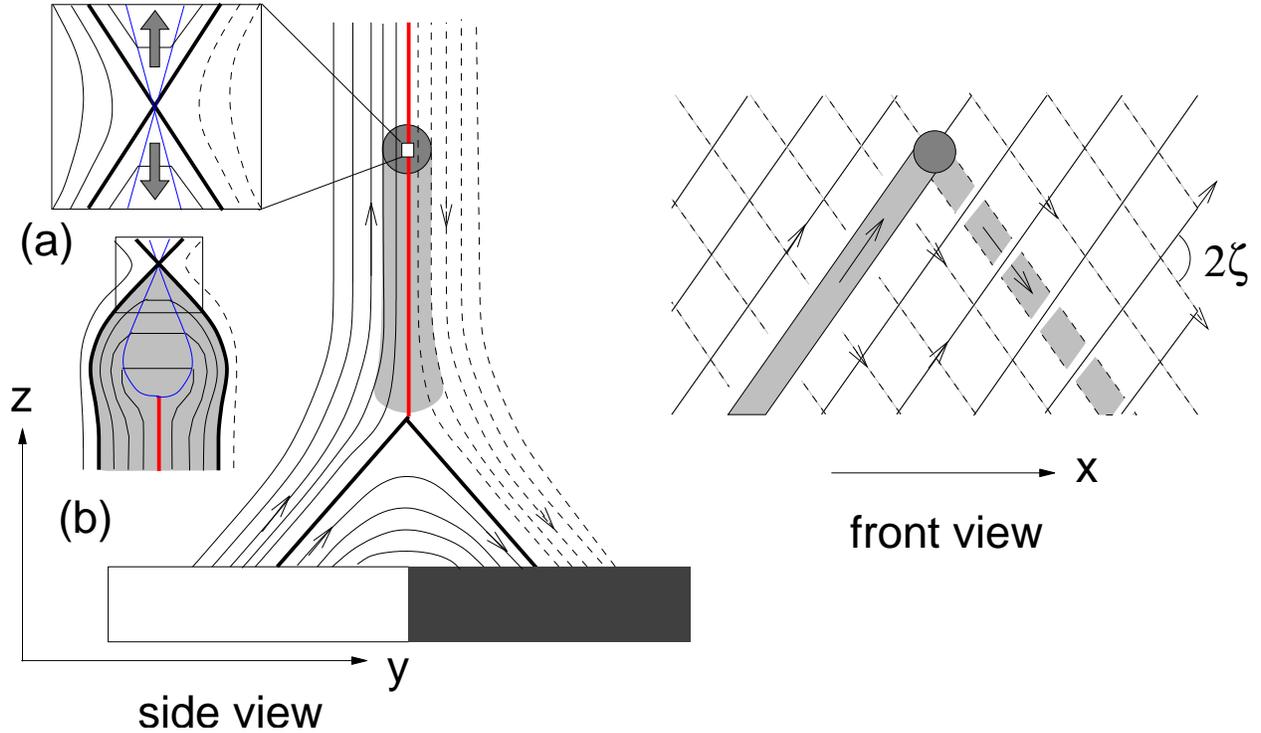}
\caption{The classical Kopp-Pneuman model and its modifications.
The side view (left) shows the basic two-dimensional geometry.  Inset
(a) on the upper left shows how steady Petschek reconnection is configured
within the current sheet.  Inset (b) on the lower left shows how a
transient reconnection produces a finite layer of reconnected flux
(grey).  The front view (right) shows how the model becomes
three-dimensional when the reconnection is localized in third
dimension ($x$).  Instead of a current layer the reconnection
produces a $\Lambda$-shaped flux tube of closed field (grey);
a completely detached, V-shaped flux tube (not shown) is
created at the same time.}
	\label{fig:geom}
\end{figure}

Magnetic reconnection has long been proposed as a mechanism for heating
coronal plasma.  In one early
model \citep[see \fig\ \ref{fig:geom}]{Kopp1976},
reconnection occurs between open field lines separated by 
a vertical current sheet (red line) creating 
new closed field lines (post-flare loops, grey).  
Closing these field lines stops the solar wind upflow in a
gas-dynamic shock (GDS) that \citet{Kopp1976} estimated would raise 
the temperature by $80\%$.  \cite{Cargill1982} found the direct
magnetic energy conversion by reconnection 
to be a far more effective source of heating in this same model.  It
could raise the temperature of post-flare loops by up to
a factor of three to 6 MK.  Even this higher value is, however, 
insufficient to explain the $15$--$20$ MK temperatures observed at
apices of post-flare loops in Soft X-ray and Fe {\sc xxiv} EUV
emission \citep{Warren1999}.

Recent theoretical investigations have revealed that 
electric fields large enough for fast magnetic reconnection
can be self-consistently produced by a wide range of
small-scale mechanisms, provided only that they are localized
within a segment of the 
current sheet \citep{Birn2001,Biskamp2001}.  All such
processes generate reconnection flows resembling the model of
\citet[see inset a]{Petschek1964}
with slow magnetosonic shocks outside the non-ideal region
(SMSs, blue lines) at which the magnetic field is deflected and
weakened, thereby heating the plasma.
When reconnection is localized in both space and time
the SMSs close back together across a finite layer of horizontal field, as
shown in inset (b) of
\fig\ \ref{fig:geom} \citep{Semenov1983,Biernat1987}.  This field forms
the top of a ``hairpin'' comprising all the
flux closed since the onset of reconnection.

As the hairpin flux sheet retracts its field lines become much
shorter and releases substantial magnetic energy.
This magnetic energy is converted almost entirely into kinetic energy
rather than partly into heat as in Petschek's steady state model
\citep{Semenov1998}.  In a strictly two-dimensional version of the 
model the mass of the shortened tube accumulates in the tip of the the
retracting hairpin.  This ``snowplowing'' artifact is absent when
reconnection is also localized in the third dimension ($x$) and 
there is a magnetic
field component in that direction.  With the horizontal field
component (sometimes called a ``guide field'') 
the current sheet separates field with angle $2\zeta$
($\tan\zeta=B_z/B_x$, see the ``front view'' of \fig\
\ref{fig:geom}).

Flux reconnected within a patch and over a finite interval 
forms a $\Lambda$-shaped flux tube (grey) similar to
magnetospheric {\em flux transfer events} 
\citep{Russell1978,Lee1993,Otto1995}.
The tube is distinguished from the surrounding field (the flux layers)
by the distinct connectivity given it through reconnection.  Different
field line geometry, such as the bend, produce dynamics
in the tube entirely different from those in the surrounding flux
layers, with which it has little subsequent interaction.

Recently \citet{Linton2006} studied the relaxation of this
post-reconnection flux tube using three-dimensional
magnetohydrodynamic (MHD) simulation.  They found that the
perpendicular dynamics of the tube, as it moves between the flux layers,
is well approximated by the equations
for a thin magnetic flux tube \citep{Spruit1981}, although with
$\beta<1$.  They presented an analytic solution in which
the post-reconnection tube shortens at the Alfv\'en
speed, converting magnetic energy into kinetic energy.
\citet{Linton2006} did not investigate the
dynamics parallel to the tube or any associated thermal
effects of shortening.  

This letter demonstrates that flux tube shortening is a powerful, inevitable 
mechanism for heating post-flare loops, which has not been previously
investigated.  The magnetic forces responsible for shortening, also drive
compressive parallel flows at the Alfv\'en speed.  At very low
$\beta$, these are high-Mach number flows 
whose collision naturally generates very strong shocks.  
The shocks are distinct from the SMSs of Petschek reconnection and are driven
by reconnection-initiated, perpendicular dynamics. This is in contrast to
previous investigations of flux tube shocks wherein acoustic
wave-steepening or pressure differences were considered as drivers 
\citep{Herbold1985,Thomas1991}. 

\section{Post-reconnection flux tube dynamics}

We begin by assuming that localized, transient, fast
magnetic reconnection has occurred, by an unspecified 
physical mechanism, within an otherwise
static current sheet.  This will, as just discussed, leave
a $\Lambda$-shaped flux tube, initially at rest.  Due to its 
sharp bend it is out of equilibrium, and
magnetic forces start it 
sliding downward between the magnetic layers separated by the current
sheet.

The tube's retraction, unhindered by the external flux layers, 
can be modeled 
using the {\em thin flux tube} equations of \citet{Spruit1981} and
subsequent authors \citep[see, for example, the review
by][]{Fisher2000}.  While the high-$\beta$ flux tubes in those
previous investigations are confined by the pressure of the
unmagnetized convection zone, our post-reconnection tube has very low $\beta$
and is confined by the magnetic pressure of the
flux layers outside the current sheet.  
The tube is an isolated entity
distinguished from its surroundings by its connectivity
\citep{Linton2006,Linton2008}.
We assume sufficient collisionality to justify the use of MHD equations 
throughout.

The tube is assumed thin enough to be described only by
its axis.  Internal properties such as the
magnetic field strength, $B_i$, pressure, $p_i$, and mass density
$\rho_i$, are function only of axial position.
The tube is also thin enough for fast magnetosonic waves to
establish pressure balance across its diameter virtually
instantaneously.  This assumption constrains the internal properties
to match those outside the flux tube: 
$B_i^2/8\pi+p_i=B_e^2/8\pi+p_e$, assumed to be uniform and constant.

We will also assume the plasma $\beta$ to be always small.  The main
force on a section of tube is therefore the magnetic tension due to
curvature of the axis $B_i^2(\partial\that/\partial\ell)/4\pi$,
where $\ell$ is arc-length
and $\that=\partial\xvec/\partial\ell$ is the unit tangent vector. 
Since it is ultimately the Lorentz force, it is natural that this
force is strictly perpendicular to the axis
($\that\cdot\partial\that/\partial\ell=0$).

The pressure gradient, $-\that\,\partial p_i/\partial\ell$, is
formally smaller, by a factor of $\beta$, than the magnetic tension.
It is, however, the only force parallel to the axis, and is
essential to arresting internal pile-up of mass.  We therefore retain
terms involving pressure to first order in
$\beta$.

Velocity evolution is governed by a momentum equation with only
these two forces.  Conservation properties become apparent
when arc-length is replaced with $\mu$, 
the integrated mass per unit flux:
$\partial\mu/\partial\ell=\rho_i/B_i$.  The value of $\mu$ never
changes for a given fluid element.  The resulting momentum equation
\be
  {d\vvec\over dt} ~=~
  {B_i\over 4\pi\rho_i}{\partial\over\partial\ell}(B_i\that)
  ~=~ {\partial\over\partial\mu}\left(\, {B_i\that\over4\pi } \right) ~~,
	\label{eq:FTe}
\ee
includes the parallel pressure gradient after use of pressure balance,
$B_i=B_e + 4\pi(p_e-p_i)/B_e$, valid to first order in $\beta$.
The momentum per unit flux of any section of tube,
\be
  {\momvec} ~=~ \int{\rho_i\vvec\over B_i}\,d\ell ~=~ \int\vvec\,d\mu
  ~~,
	\label{eq:mom}
\ee
changes only through forces (per unit flux)
from the ends of the section, $B_i\that/4\pi$,
directed parallel to the axis.

\section{Shock relations in thin flux tubes}

Momentum conservation leads to a set of shock relations for thin flux
tubes.  Consider two straight sections with uniform
properties (designated $1$ and $2$), 
separated by an abrupt change at coordinate $\mu_0$.
The length-scale of this change is large compared 
to the tube radius but otherwise small
enough that we hereafter call it a
``discontinuity'' and ``corner''.\footnote{Analysis of the
discontinuity's internal structure must be done outside the thin-tube
approximation.  Doing so will probably reveal its length-scale to be
comparable to the tube's radius.
This is analogous to non-ideal effects resolving internal structure
of a hydrodynamic shock at scales comparable to the mean free path
\citep{Grad1951}. Following that analogy, the thin flux tube equations
provide external conservation laws leading to the shock relations.}
This feature moves through space at
constant velocity $\uvec$ while its Lagrangian coordinate changes at
constant rate $\dot{\mu}_0$. We assume that the properties of the 
straight sections do not change as this
happens.  Fluid velocities on either side of
the discontinuity differ from $\uvec$, but
components perpendicular to the tangent vector match that
of $\uvec$: $(\vvec-\uvec)\times\that=0$.

The component of relative velocity parallel to the tangent vector,
$\vpar=\that\cdot(\vvec-\uvec)$, represents a flow across the
discontinuity.  The mass flux (per magnetic flux) through the
discontinuity, $\dot{\mu}_0$, must equal the mass flux across points
on either side of it, since the corner moves
without changing.  For positions $\mu_1$ and $\mu_2$, separated by
fixed distances from $\mu_0$ this means
\be
  \dot{\mu}_2 = -\left.{\rho_i\vpar/B_i}\right\vert_2
  = -\left.{\rho_i\vpar/B_i}\right\vert_1 ~=~\dot{\mu}_1 
  ~=~\dot{\mu}_0~~.
	\label{eq:mass_cons}
\ee

The flux tube between $\mu_2$ and $\mu_1$ does not change so its
momentum (per magnetic flux) is constant,
\be
  \dot{\momvec} = \disc{\dot{\mu}\vvec} + \disc{B_i\that/4\pi}
  = 0 ~~,
\ee
denoting $\disc{f} = f_2 - f_1$.  Since
$\vvec=\uvec+\vpar\that$ away from the discontinuity
momentum constancy may be written
\be
  \avg{B_i/4\pi - \rho_i\vpar^2/B_i}\disc{\that} 
  -\disc{p_i/B_e + \rho_i\vpar^2/B_i}\avg{\that} =0 ~~,
	\label{eq:mom_cons}
\ee
where $\avg{f}=(f_2+f_1)/2$.

The vectors $\disc{\that}$ and $\avg{\that}$ are orthogonal
so \eq\ (\ref{eq:mom_cons}) represents two independent 
equations.  As long as the bend is not a hairpin
($\avg{\that}\ne0$) the $\avg{\that}$
component of momentum conservation, to lowest order in $\beta$, is 
equivalent to gas-dynamic momentum conservation: 
$\disc{p_i +\rho_i\vpar^2}=0$.  To the same order in $\beta$, \eq\
(\ref{eq:mass_cons}) gives the gas-dynamic mass continuity equation:
$\disc{\rho_i\vpar}=0$.  Provided there is no heat flow from the
external field or across $\mu_1$ or $\mu_2$, then the sum of kinetic
and thermal energy within the tube section will also be conserved.
This provides one additional, independent constraint on
$\rho_i$, $p_i$ and $\vpar$ across the discontinuity.
(Since $B_i$ can be related directly to $p_i$ it can be
eliminated from energy conservation.)

The foregoing describes three relations between six
quantities which do not include the magnetic
field's strength or direction.  The
relations are the traditional gas-dynamic Rankine-Hugoniot
conditions \citep{Courant1948}, but for flows inside a flux tube 
\citep{Ferriz-Mas1987}.  The other component of \eq\ (\ref{eq:mom_cons}), to
lowest order in $\beta$,
\be
  \avg{1 - (4\pi\rho_i/B_e^2)\vpar^2}\,
  \bigl\vert\disc{\that}\bigr\vert^2 = 0 ~~.
	\label{eq:bend_eq}
\ee
constitutes one more relation, which does involve the field direction.

From values on one side the Rankine-Hugoniot relations may be
satisfied in two different ways by values on the other \citep[see for
example][]{Courant1948}.  They may be satisfied non-trivially by a unique
set of {\em different} values, or they may be satisfied trivially by
the {\em same} set of values
(i.e.\ $\disc{\rho_i}=\disc{\vpar}=\disc{p_i}=0$).
In the non-trivial case, the three different
quantities satisfy three independent constraints and cannot be forced,
in general, to satisfy a fourth.  The leading factor of \eq\
(\ref{eq:bend_eq}), however, constitutes a fourth independent
constraint so it will not in general vanish; it is therefore
necessary that $\disc{\that}=0$ (there is no bend).
This means that a thin, low-$\beta$ flux
tube can support discontinuities in internal quantities
only at a GDS within a straight section of tube.

If, on the other hand, there
{\em is} a bend in the flux tube ($\disc{\that}\ne0$), the only
way to satisfy \eq\ (\ref{eq:bend_eq}) is for the
pre-factor to vanish.  Since the non-trivial solution of all three 
Rankine-Hugoniot conditions would over-determine the system, they must
be satisfied trivially, without discontinuity.  In other words
$\rho_i$, and $\vpar$ are continuous at $\mu_0$ and satisfy the
relation $\vpar=B_e/\sqrt{4\pi\rho_i}=\va$ the Alfv\'en speed.  This
is similar to an intermediate shock \citep{Priest2000}, but includes the
influence of fast magnetosonic waves assumed to be maintaining pressure
balance across the tube.

\section{Shocks in the retracting flux tube}

\begin{figure}[htp]
\epsscale{0.95}
\plotone{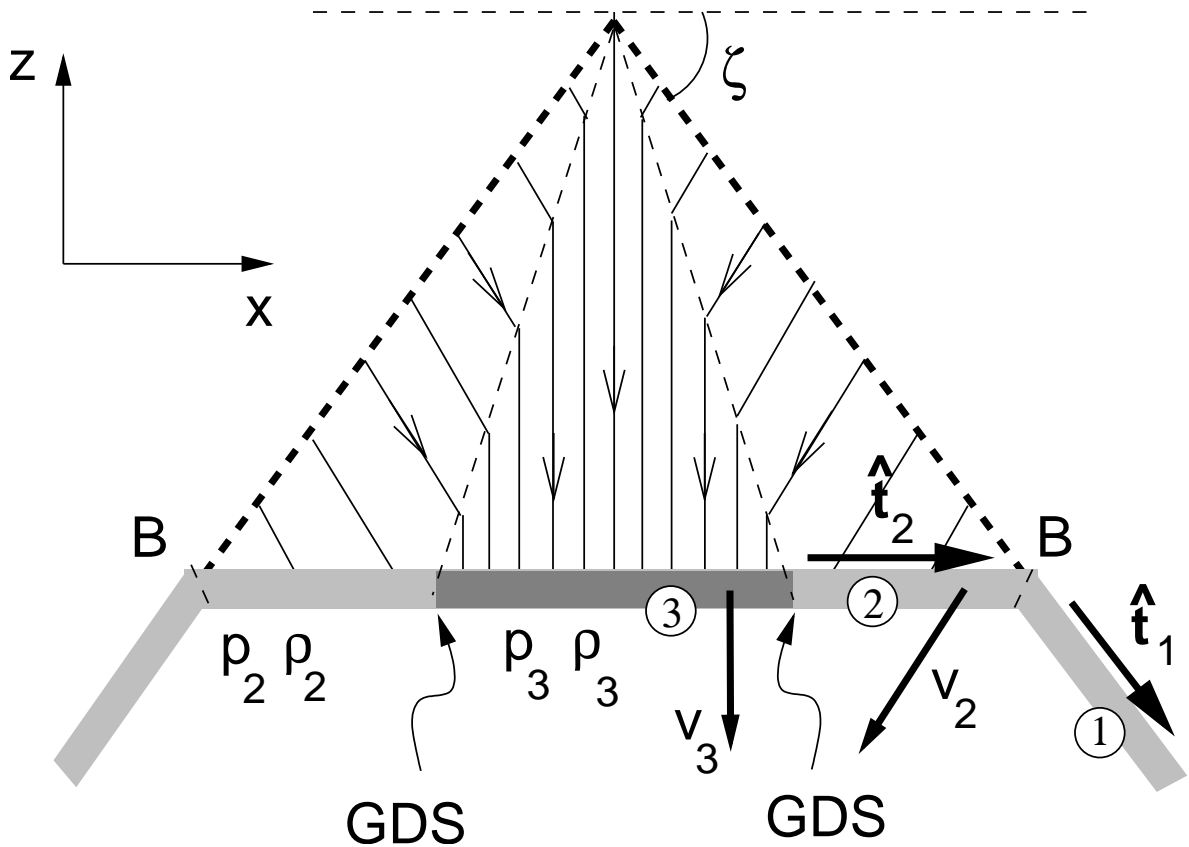}
\caption{The flux tube retracting from the $\Lambda$-shaped initial
condition shown in \fig\ \ref{fig:geom}.  This initial configuration is
shown by thick dashed lines, while its current state is shown in
grey.  The past trajectory of the fluid, shown by solid lines, is
initially focused inward.  The gas-dynamic shocks (GDS) 
redirect the flow vertically; their past positions are shown by a thin
dashed line.  Circled numbers designate the
states, $1$, $2$ and $3$, separated by these shocks.}
	\label{fig:bend}
\end{figure}

The $\Lambda$-shaped bend in the post-reconnection flux tube shown in
\fig\ \ref{fig:geom} will immediately decompose into four different
shocks of the kinds described above.  Two intermediate shocks (bends,
B) propagate along the field lines, forming a straight horizontal
section between them (see \fig\ \ref{fig:bend}), as previously found
by \citet{Linton2006}.
Two GDSs propagate away
from the center, at $\pm v_s$, along the horizontal section.
This symmetric arrangement divides the
tube into sections labeled, $3$, $2$ and $1$, outward from the center.

Section $1$ consists of the flux tube at rest and in its initial state:
$\vvec_1=0$, $\rho_1=\rho_e$ and $p_1=p_e$.  The initial field lines
are inclined at angle $\zeta$, so 
$\that_1=\xhat\cos\zeta-\zhat\sin\zeta$, while
$\that_2=\xhat$ in the horizontal section.  The bend
propagates along the initial field at the Alfv\'en speed:
$\uvec=\vae\that_1$.  The relative parallel flow is continuous
across the bend
$v_{\parallel,1}=v_{\parallel,2}=-\vae$.  These facts can be combined
into the post-bend fluid velocity
\be
  \vvec_2 = \uvec + \vpar\that_2 = 
  -2\vae\sin^2(\zeta/2)\xhat-\vae\sin\zeta\zhat ~~,
\ee
directed along the bisector of the bend 
(the mean direction of the curvature force).

Internal quantities are continuous across the bends, so
$\rho_2=\rho_1=\rho_e$ and $p_2=p_1=p_e$.
From a reference frame moving downward at $-\zhat\vae\sin\zeta$, sections
$2$ and $3$ appear to form a classic shock tube, with inflow at
\be
  M_2 = {|v_{x,2}|\over c_{s,e}} = 
  \sqrt{8\over\gamma\beta_e}\sin^2(\zeta/2) ~~,
\ee
where $\gamma=5/3$ is the ratio of specific heats.
Since $\beta_e\ll1$, this Alfv\'enic inflow can have extremely high
Mach number when field lines of significantly different 
orientation reconnect.

This supersonic inflow is brought to rest, $v_{x,3}=0$, by a GDS
moving outward at $v_s$.  This stopping shock, equivalent to a piston
moving into stationary fluid at speed $|v_{x,2}|$, is a classic
problem \citep[see \S 69 from][for example]{Courant1948}, for which the
solution is
\be
  {v_s/|v_{x,2}|} = \sqrt{M_2^{-2}+(\gamma+1)^2/16}
  - \quarter(3-\gamma) ~~.
\ee
The density ratio, $\rho_3/\rho_2=1+|v_{x,2}|/v_s$, following from mass
conservation, approaches the well-known limit
$\rho_3/\rho_e=(\gamma+1)/(\gamma-1)=4$, at large Mach numbers.

The post-shock pressure,
$\beta_3 = \beta_e[1+\gamma M_2^2(1+v_s/|v_{x,2}|)]$,
also follows
from the shock relations, and
even in the limit of vanishing pre-shock pressure ($\beta_e\ll1$), 
it can be significant:
$\beta_3\simeq4(\gamma+1)\sin^4(\zeta/2)$.
The low-$\beta$ assumption thereby imposes a limit on the reconnection angle; 
only for $\zeta<36^{\circ}$ is $\beta_3<0.1$ and 
$\beta_3>1$ when $\zeta>67^{\circ}$.

\begin{figure}[htb]
\epsscale{0.75}
\plotone{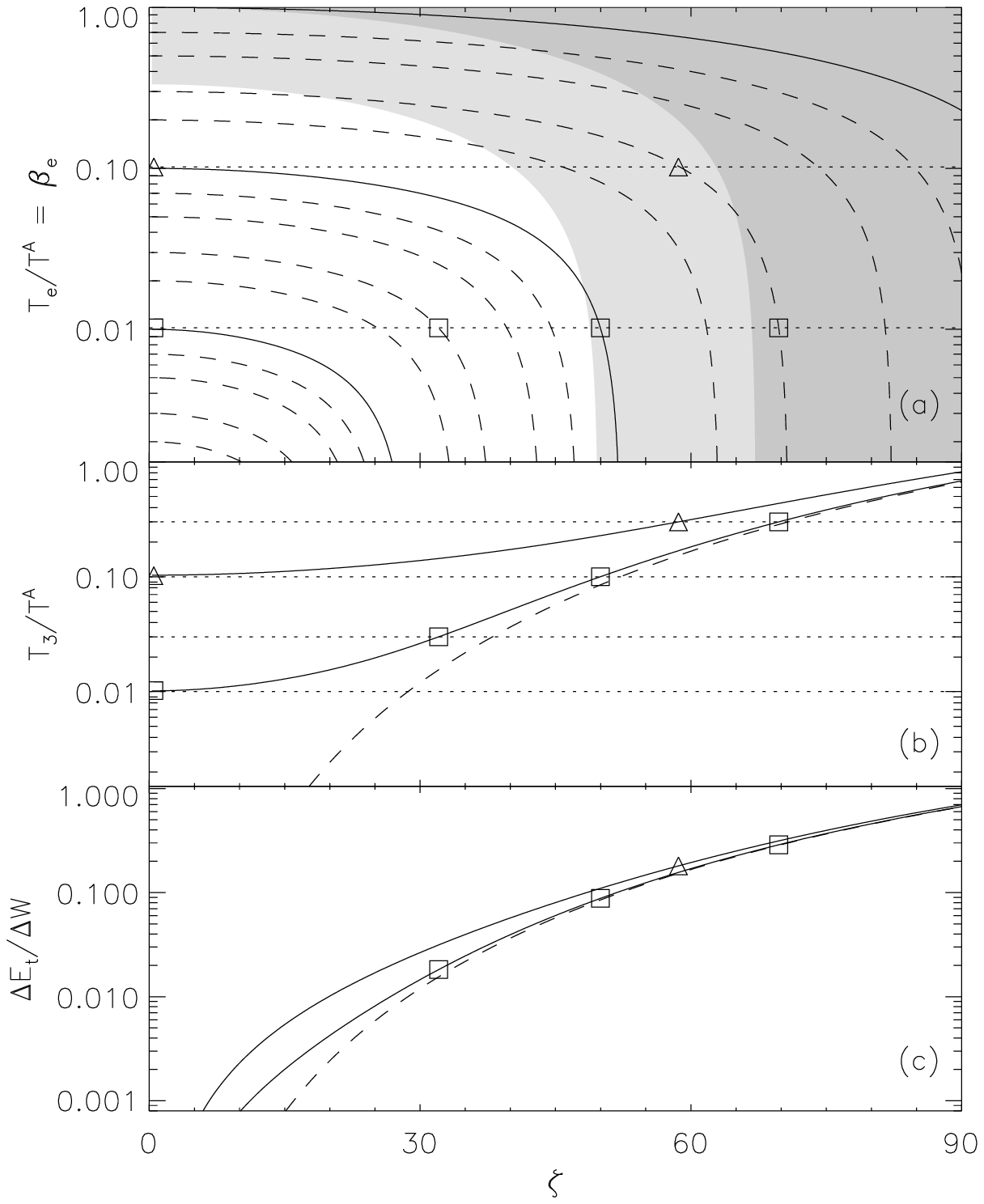}
\caption{The post-shock temperature and thermal energy across the
$\zeta$--$\beta_e$ parameter space.  The top panel (a) shows contours
of $T_3$ whose levels can be determined from intersection with the 
left axis ($\zeta=0$) since  $T_3=T_e$ there.
Solid contours show  $T_3=0.01,\, 0.1$ and 1 (all in units of $\ta$).
Light and dark grey regions show $0.333<\beta_3<1$ and $\beta_3>1$
respectively.  The middle panel (b) shows $T_3$ versus $\zeta$ 
for the values $T_e=\beta_e=0.01$ (squares) and $0.1$ (triangles).
The bottom dashed curve is limit for the case $\beta_e=0$.
The bottom panel (c) shows, for the same values of $T_e=\beta_e$, 
the fraction of released energy thermalized
by the GDS.  Symbols show the same point in each panel.}
	\label{fig:temp}
\end{figure}

The post-shock temperature is more conveniently 
expressed with respect to 
$\ta=(m_p/k_B)\vae^2/2$, than to
the initial temperature.
For example, $15$ G field immersed in 
$n_e=3\times10^8\,{\rm cm}^{-3}$ plasma has a characteristic
temperature $\ta=2\times10^8$ K.
The plasma beta is $\beta_e = T_e/\ta$, so a
2 MK coronal plasma will have $\beta_e=10^{-2}$.
Figure
\ref{fig:temp}a--b shows the post-shock temperature,
$T_3=\ta\beta_3\rho_e/\rho_3$, over a range of
$\beta_e$--$\zeta$ parameter space.  For $\beta_e=0.01$ and
$\beta_3=1/3$ (the edge of the light grey area, 
$\zeta\simeq50^{\circ}$) the 
post-shock plasma will be $T_3=0.1\ta=20$ MK.

The total energy release initiated by reconnection,
$\Delta W=2\vae^2L\rho_e\sin^2(\zeta/2)$, 
is the initial kinetic energy of all 
mass within the $L=2\vae t$ of tube affected by retraction.  This 
is greater than the energy decrease resulting from 
shortening the tube by $\Delta L=2L\sin^2(\zeta/2)$, due to additional
work done by the background magnetic field expanding into the vacated
volume.  In our idealized model this work exactly doubles the energy,
but in more realistic scenarios the factor may be somewhat
different.

The moving mass is deflected downward by the
GDS, converting a portion of its kinetic
energy into thermal energy.  The kinetic energy (per area) 
converted to thermal energy, $\rho_3v_{x,2}^2v_st$, constitutes a
fraction
\be
  {\Delta E_t\over\Delta W}=
  {\rho_3v_{x,2}^2v_s\over\rho_e|\vvec_2|^2\vae} =
  2\left(1+{v_s\over|v_{x,2}|}\right)\sin^4(\zeta/2) ~~,
	\label{eq:ET}
\ee
of the total (see \fig\ \ref{fig:temp}c).
Cases of vanishing initial pressure ($\beta_e\ll1$) will have a
fraction $(\gamma+1)\sin^4(\zeta/2)$ of the released energy
thermalized.  Compression work done on larger initial pressure will
raise this fraction slightly, but all cases with $\beta_3<1/3$ 
thermalize less than $10\%$ of the released energy.

\section{Summary}

The foregoing illustrates, through a simplified analytic model, 
a process
we believe must be common in the flaring corona.  Localized reconnection
does not directly dissipate magnetic energy, but rather initiates its
release through subsequent shortening of field lines.  This shortening
propagates from the reconnection site at the Alfv\'en speed,
converting energy from magnetic into kinetic form.  The shortening
flux tubes compress plasma within them, raising its temperature as
they do so.  Due to the supersonic (Alfv\'enic) flows this compression
occurs at strong shocks, which raise the temperature far beyond that from
adiabatic compression.

Both the GDS in our model and the SMS in Petshcek's model
result, ultimately, from shortening or weakening of magnetic field
lines.  The two-dimensional Petschek model permits shortening only
perpendicular to the symmetry direction, accompanied 
by weakening.  The shock is therefore a SMS whose normal is mostly
perpendicular to the sheet.  After three-dimensional patchy
reconnection, on the other hand, field lines also shorten in the
erstwhile symmetry direction which is the orientation of the shock
normal ($\pm\xhat$).
The GDSs are disconnected from the diffusion
region, and are instead features of the ideal relaxation {\em following}
reconnection.  
This new scenario requires two-step energy conversion, from magnetic
to kinetic to thermal energy, in contrast to the SMSs which
thermalize magnetic energy directly where they weaken the field.  

\medskip

This work was supported by NASA and the NSF.


\end{document}